\documentclass[prd,letterpaper,onecolumn,amsmath,amsfonts,amssymb,nofootinbib]{revtex4}
\pdfoutput=1
\usepackage {amsmath}
\usepackage[dvips]{graphicx}
\usepackage{feynmp}
\usepackage{amssymb}
\newcommand{\be}{\begin{equation}}
\newcommand{\ee}{\end{equation}}
\newcommand{\bear}{\begin{eqnarray}}
\newcommand{\eear}{\end{eqnarray}}

\newcommand{\lapproxeq}{\lower .7ex\hbox{$\;\stackrel{\textstyle
<}{\sim}\;$}}
\newcommand{\gapproxeq}{\lower .7ex\hbox{$\;\stackrel{\textstyle
>}{\sim}\;$}}
\newcommand{\stackdown}[2]{\lower 1.4ex\hbox{$\;\stackrel{\textstyle{#1}}
{\scriptstyle{#2}}\;$}}
\newcommand{\beq}{\begin{equation}}
\newcommand{\eeq}{\end{equation}}

\newcommand{\ba}{\begin{eqnarray}}
\newcommand{\ea}{\end{eqnarray}}

\newcommand{\bea}{\begin{eqnarray}}
\newcommand{\eea}{\end{eqnarray}}

\makeatletter
\def\slash{\@ifnextchar[{\fmsl@sh}{\fmsl@sh[0mu]}}
\def\fmsl@sh[#1]#2{%
  \mathchoice
    {\@fmsl@sh\displaystyle{#1}{#2}}%
    {\@fmsl@sh\textstyle{#1}{#2}}%
    {\@fmsl@sh\scriptstyle{#1}{#2}}%
    {\@fmsl@sh\scriptscriptstyle{#1}{#2}}}
\def\@fmsl@sh#1#2#3{\m@th\ooalign{$\hfil#1\mkern#2/\hfil$\crcr$#1#3$}}
\makeatother
\usepackage{color}

\definecolor{orange}{rgb}{0.9,0.2,0}
\definecolor{brown}{rgb}{0.7,0.3,0.2}
\definecolor{fuxia}{rgb}{1,0,1}
\definecolor{skyblue}{rgb}{0,0.1,0.9}
\definecolor{violetred}{rgb}{0.8,0.13,0.56}
\definecolor{deeppink}{rgb}{1.00,0.08,0.5}
\definecolor{pink}{rgb}{1.00,0.75,0.80}
\definecolor{orchid}{rgb}{0.85,0.44,0.84}
\definecolor{lightpink}{rgb}{1.00,0.71,0.76}
\definecolor{bluish}{rgb}{0,0.6,0.8}
\begin{document}
\title{Inflation in $R^2$ supergravity with non-minimal  superpotentials }
\author{G. A.~\ Diamandis, B. C.~\ Georgalas,  K.~\ Kaskavelis, P.~\ Kouroumalou,} 
\author{A. B.~\ Lahanas}  \email{alahanas@phys.uoa.gr} \author{G.~\ Pavlopoulos}
\affiliation{University of Athens, Physics Department,  
Nuclear and Particle Physics Section,  
GR--15771  Athens, Greece}

\begin{abstract}
We investigate the cosmological inflation in a class of supergravity models that are generalizations of non-supersymmetric $R^2$ models. Although such models have been extensively studied recently, especially after the launch of the PLANCK and BICEP2 data, the class of models that can be constructed has not been exhausted. In this note, working in a supergravity model that is a generalization of Cecotti's model, we show that  the appearance of new superpotential terms, which are quadratic in the superfield $\, \Lambda$  that couples to the Ricci  supermultiplet, alters substantially the form of the scalar potential. 
The arising potential has the form of the Starobinsky potential times a factor that is  exponential in the inflaton field and dominates for large inflaton values.  We show that the well-known Starobinsky inflation scenario is maintained only for unnaturally small fine-tuned values of the coupling describing the  $\Lambda^2$ superpotential  terms.  A welcome feature is the possible increase of the tensor to scalar ratio $r$, within the limits set by the new Planck and BICEP2 data.   
\end{abstract}
\maketitle

\vspace*{1.0cm}
\section{Introduction} 

Models of inflation are constrained by  observations of WMAP \cite{WMAP}  and Planck \cite{PLANCK} satellites. 
The spectral index is found in the range $\, n_s = 0.9608 \pm 0.0054$ while the tensor to scalar ratio is bounded from above 
$\, r < 0.111$.  In addition  the BICEP2 \cite{BICEP}  experiment, claiming  for  the discovery of primordial gravitational waves resulting to a ratio $\, r =0.16^{+0.06}_{-0.05}$,  aroused the interest of both experimentalists and theorists.    
PLANCK satellite data \cite{PLANCK} are in perfect agreement with the Starobinsky model of inflation \cite{STAR} which predicts 
a tensor to scalar ratio in the range $\, r \simeq 0.004$, which is almost two orders of magnitude smaller than the claimed discovery of BICEP2 which  points  towards  chaotic inflation \cite{Linde:1983gd} .
In the meantime Planck collaboration released new data with increased precision \cite{PLANCK2}, which are in agreement with the previous data,  according to which $\, n_s = 0.968 \pm 0.006$ and $\, r < 0.11$. Also BICEP2 and Planck joined collaboration \cite{Ade:2015tva}   established a robust upper bound  $ r < 0.12 $, which is substantially lower than the value quoted in \cite{BICEP}. 

Much effort  has been expended towards building inflationary models embedded in the framework of supergravity theories.  Chaotic inflation \cite{Linde:1983gd} scenario can be incorporated in supergravity schemes \cite{ Goncharov:1983mw,YANA} and more recently general chaotic inflationary supergravity potentials have been studied  \cite{SUGRAchaos}. 
Supergravity models that incorporate $R + R^2$ terms and reproduce Starobinsky's  inflation predictions for $\, r , n_s$ have received a lot of attention recently\cite{ALVAREZ,su1,su2,su3,su4,su5,su6,FERRARA,FERPRO,MAVRO1,MAVRO2,KAL3,su8,su7,su66,su9,su10,su11,su12,FERRARA2,su13,KETOV2a,FER3,KETOV2b,NANOP}. A class of  supergravity models are  described by  no-scale K\"ahler potentials \cite{lahanas} 
and many of the proposed inflationary models have a no-scale structure 
\cite{su1,su2,su6,su7,su11,su12,KETOV2a,NANOP}. It is worth noting that 
in this class of models there is the possibility of accommodating models interpolating between low ( $r \sim 0.001$) and large values 
( $r \sim 0.1$) depending on the parameters.  This can be also accomplished in attractor solutions that relate in a continuous manner  the predictions of the Starobinsky model with those of the quadratic chaotic potential \cite{ATTRACTORS,ATT2}.

Among the possible theoretical schemes, incorporating  the virtues of the Starobinsky $R^2$ model that lead to successful inflation, are higher derivative supergravity Lagrangians \cite{Theisen,cecotti,CEC2,HIND,DALIA}. In these, besides the matter chiral and vector multiplets, additional chiral multiplets are unavoidably introduced. In the minimal scenario \cite{cecotti} one uses two multiplets that after  eliminations of the auxiliary fields involved leads to a supergravity Lagrangian including $R^2$. 
This extends the  Starobinsky model in a non-trivial manner in the sense that additional terms appear, in comparison with the non-supersymmetric theory,  that only conditionally can sustain a successful  inflationary scenario.

Our aim in this note is to further investigate natural extensions of some of these models that include superpotential couplings not considered in previous works. Such couplings lead to inflaton potentials that do not have the Starobinsky form, unless some of the parameters are fined tuned. The question that arises is under what conditions these supergravity  generalizations lead to successful inflation and under which circumstances 
large values of the ratio $r$ can be obtained saturating the upper bounds set on $r$ by Planck and BICEP2. 

This article is organized as follows : \\
In section 2 we review the equivalence of $R^2$ supergravity to an ordinary supergravity theory, by generalizing the original Cecotti's Lagrangian \cite{cecotti}, working in the framework of the Poincare supergravity. 
In section 3, working in the context of this framework,  we discuss  the properties of a class of models   as far as cosmological inflation is concerned. Section 4 is devoted to a numerical treatment of the cosmological solutions and the consequences for the parameters describing the model under consideration. In section 5 we discuss the cosmological predictions paying special emphasis on the tensor to scale ratio $r$. We end  up with the conclusions.

\section{Reviewing the $R^2$ supergravity model}
It has been proven that  $R^2$ gravity \cite{whitt}, upon supersymmetrization is equivalent to ordinary supergravity theory which includes  two chiral multiplets with specific couplings \cite{cecotti} .  In this note we start by  considering the locally supersymmetric Lagrangian of SuperPoincare algebra which involves two chiral multiplets $\Phi,\, \Lambda$ allowing for non-linear superpotential couplings of the field $\Lambda$ that couples to the curvature multiplet.  One can also start from a superconformal action, see for example \cite{FERPRO, KAL3}, and then proceed to  Poincare supergravity action by a proper gauge fixing but this case is not considered  in this work. 

Our starting Lagrangian is therefore given by 
\bea
\mathcal{L} \,=\, \int d^2\Theta 2 \mathcal{E} \left[  - \Lambda  \mathcal{R} -\frac{1}{8} \left(
\bar{\mathcal{D}}\bar{\mathcal{D}} - 8 \mathcal{R} \right) \Omega (\Phi , \bar{\Phi }) + W(\Phi, \Lambda  ) \right] + h.c.
\label{lag1}
 \eea
which is described by the real kinetic function $  \Omega (\Phi , \bar{\Phi }) $, which is a function of $ \Phi, \bar{\Phi} $,  and a superpotential $W(\Phi, \Lambda)$  which is an holomorphic function of the supermultiplets $\Phi,\, \Lambda$  .
Note that the curvature supermultiplet  $  \cal{R}$  couples to a chiral multiplet  which we denote  by $\Lambda$ and this choice covers the most general case up to field redefinitions. 
The particular case $W(\Phi, \Lambda) = \Lambda \, \Phi $, in which  $\Phi$ and $\Lambda$ are coupled linearly to each other,  is the one studied  by Cecotti but evidently other options are available.  
The above Lagrangian can be cast in the  form,
\bea
\mathcal{L} \,=\, \int d^2\Theta 2 \mathcal{E} \left[  -\frac{1}{8} \left(
\bar{\mathcal{D}}\bar{\mathcal{D}} - 8 \mathcal{R} \right) \, \Omega ' (\Phi , \Lambda , \bar{\Phi }) 
 + W(\Phi, \Lambda  ) \,  \right] + h.c.
 \label{lag2}
 \eea
which is also a supergravity Lagrangian with a redefined kinetic function
\[
\Omega ' (\Phi , \Lambda , \bar{\Phi }, \bar{\Lambda }) \,=\,  \Omega (\Phi , \bar{\Phi }) -\frac{1}{2}(\Lambda +\bar{\Lambda })
\, .
\]
The superpotential can be brought to   the following form by merely segregating the linear in  $\Phi$ terms, which in general couples to a superfield which is a function of the chiral multiplet $\Lambda$, 
\[
W(\Phi , \Lambda ) \,=\, g(\Lambda) \Phi + P(\Phi, \Lambda  )  \, .
\]
The function $\, g(\Lambda) $  is non-linear, in general, and its departure from linearity brings about new features that may affect the cosmological evolution as we shall see. 

For our purposes it suffices to keep only the bosonic fields  and thus the superfields can be  expanded in the following way 

\bea
\Phi \, =  \, \varphi + \Theta \Theta \, F_\varphi  \quad ,   \quad   \Lambda \, = \, \lambda + \Theta \Theta  F_\lambda 
\nonumber
\eea
leading to 
\bea
\displaystyle g(\Lambda ) \, =\, g(\lambda ) + \Theta \Theta \, \displaystyle \frac{\partial g}{\partial \lambda } \,  F_\lambda 
 \quad , \quad
W(\Lambda , \Phi ) \,=\, W(\lambda , \varphi ) + \Theta \Theta \, (W_\lambda F_\lambda \,+\, W_\varphi F_\varphi )
\, .
\nonumber
\eea
An analogous treatment for the supergravity chiral  multiplets $ \mathcal{E} $  and $ \mathcal{R}$, following standard notation,   leads to 
\[
2 \mathcal{E} = e\{1 - \Theta \Theta \bar{M} \}
\]
and
\[
\mathcal{R} = -\frac{1}{6} \left\{M + \Theta \Theta \left[ -\frac{1}{2}R + \frac{2}{3}M \bar{M} + \frac{1}{3} b_\mu b^\mu - i D_\mu b^\mu  \right] \right\}.
\]
Then the bosonic part of the off-shell Lagrangian is written as
\[
\begin{array}{ccc}
 e^{-1} \mathcal{L}_B \,&=&\, \displaystyle  \frac{1}{6} \left[ \Omega - \frac{1}{2}(\lambda +\bar{\lambda} ) \right]
\left(R + \frac{2}{3} M \bar{M} - \frac{2}{3} b_{\mu }b^{\mu }  \right) -
\Omega _{\varphi  \bar{\varphi }} \partial _{\mu } \varphi  \partial ^{\mu } \bar{\varphi }  +
\Omega _{\varphi  \bar{\varphi }}  F_{\varphi} \bar{F}_ {\bar{\varphi }} 
 \\
&{}& \\
&-& 
\displaystyle     \frac{i}{3} 
 \left[\Omega_\varphi  \partial _{\mu } \varphi - \Omega _{\bar{\varphi }} \partial _\mu  \bar{\varphi } \right]b^\mu  -
 \frac{i}{6}(\lambda - \bar{\lambda })\mathcal{D}_\mu b^\mu - \left[ (g(\lambda) \varphi  + P) \bar{M}+ h.c. \right] \\
 &{}& \\
 &+& \displaystyle \left\{ \left[g(\lambda ) + P_\varphi - \frac{1}{3} M \Omega _\varphi   \right] F_\varphi  + 
 \left[\frac{\partial g}{ \partial \lambda }\varphi  + P_\lambda  
+ \frac{1}{6} M    \right] F_\lambda  + h.c. \right\}
\, .
 \end{array}
\] 
The ordinary supergravity model is derived by solving the equations of the auxiliary fields $M, b_\mu , F_\lambda , F_\varphi $ in the usual manner. However in this frame ( Jordan frame )   the field $\lambda $ has no kinetic term so alternatively one can use its equation of motion in eliminating the auxiliary fields. 
In doing this the equation of motion for the field $\, \lambda$ yields 
\[
\begin{array}{c}
 \displaystyle \left[\frac{\partial g}{\partial \lambda } \varphi + P_{\lambda } \right] \bar{M} +\frac{1}{12} \left[ R + \frac{2}{3}M\bar{M} - \frac{2}{3}b_\mu b^\mu  + 2i D_\mu b^\mu \right] \\
 {}\\
- \displaystyle \left(\frac{\partial g}{\partial \lambda } + P_{\varphi \lambda } \right)F_\varphi  -  \left( \frac{\partial ^2 g}{\partial  \lambda ^2}\varphi + P_{\lambda \lambda }  \right)F_\lambda  = 0
\end{array}
\]
while those stemming from $\,  F_\lambda $ and $\, M$ yield respectively
\[
\frac{\partial g}{\partial \lambda } \varphi + P_{\lambda } + \frac{1}{6} M  = 0 
\]
and 
\[
\displaystyle  \frac{1}{3} \Omega ' \bar{M} - \Omega _\varphi  F_\varphi  + \frac{1}{2} F_\lambda - 3 (\bar{g} \bar{\varphi } + \bar{P}) = 0 
\, .
\]
Eliminating $M, F_\lambda $  and substituting  back in the Lagrangian the resulting expression for $F_\varphi $  we get a  supergravity model which includes $R^2$ term. 
\be
\frac{1}{144} \dfrac{\Omega _{\varphi  \bar{\varphi }}}{\left| \dfrac{\partial g}{\partial \lambda } + P_{\varphi \lambda} + 2 \Omega _\varphi  
\left( \dfrac{\partial ^2 g}{\partial  \lambda ^2}\varphi + P_{\lambda \lambda}  \right)  \right|^2} \;  R^2
\, .
\label{r2}
\ee
Note that this term arises from the $F_\varphi  \bar{F}_{\bar{\varphi }}$ term of the original off-shell Lagrangian 
{\footnote{In the case of non-linear $g(\lambda )$  the $R^2$ coefficient is different from the one appearing in \cite{KAL3}. The difference is due to the specific superconformal gauge fixing adopted in that work.}}. 
The prefactor  of $R^2$ in eq. (\ref{r2}), in the general case, is a complicated function of the fields involved but it simplifies a great deal, becoming actually a constant, when $\Omega(\Phi,  \bar{\Phi}) =\Phi \, \bar{\Phi}$,  $g(\Lambda) = \Lambda$ and $P(\lambda, \Phi) = 0$, retrieving  in this way  Starobinsky's model. 
More general cases regarding mainly the field $\Phi $ have also been studied in \cite{HIND} as far as the vacuum structure and the supersymmetry breaking is concerned. Recently in \cite{DALIA}, besides the vacuum structure, the inflationary properties of generalized cases have been addressed.

We have thus seen that the Lagrangian given by Eq. (\ref{lag1}) and (\ref{lag2}) yields the "dual" prescription of $R^2$ supergravity where  the $R^2$ term couples with two fields. This result can be generalized with the inclusion of additional chiral multiplets as well. The crucial point for the $R^2$ description is the existence of a field $\lambda $ for which $\Omega^{'} _{\lambda \bar{\lambda }} = 0$.  In general we expect that $R^2$ terms will naturally arise from corrections  which may involve, for instance, the dilaton and other moduli fields coming from  string theory \cite{Gross} and so identifying  the field $\lambda $ with one of these fields we actually consider a model with non-linear dependencies on this field. In this sense the study of models with nonlinear behaviour,  besides being interesting  per se, it is useful in order to investigate  the particular role  these fields may play in the cosmological evolution. 
 
\section{No - scale supergravity with $\Lambda^2$ terms}
As  already outlined in the previous section  we consider  supergravity models described, in the Einstein frame, by the 
K\"{a}hler function 
{\footnote{
In this work we follow closely the notation of Bagger and Wess \cite{bagger} .
}}
\[
K \,=\, -3 ln \left( - \frac{\Omega '}{3}  \right)
\]
and a superpotential whose scalar component reads 
\[
W(\lambda , \varphi ) \,=\, g(\lambda ) \varphi + P(\lambda , \varphi ).
\]
For our considerations   the relevant terms are the kinetic terms of the fields $\lambda , \varphi $
\[
-\frac{1}{2} \left[ K_{\varphi  \bar{\varphi }} \partial _\mu \varphi \partial ^\mu \bar{\varphi } \,+\, 
K_{\varphi  \bar{\lambda  }} \partial _\mu \varphi \partial ^\mu \bar{\lambda  } \,+\,
K_{\lambda  \bar{\varphi }} \partial _\mu \lambda  \partial ^\mu \bar{\varphi } \,+\,
K_{\lambda   \bar{\lambda  }} \partial _\mu \lambda  \partial ^\mu \bar{\lambda } \right] 
\]
where
\[
K_{\varphi  \bar{\varphi }}  \,=\, \frac{3}{\Omega ^{'2}} \left( \Omega _\varphi  \Omega _{\bar{\varphi }} - \Omega' \Omega _{\varphi  \bar{\varphi }}  \right), \,\,
K_{\lambda   \bar{\lambda  }} \,=\, \frac{3}{4 \Omega ^{'2}} , \,\,
K_{\lambda   \bar{\varphi   }} \,=\, - \frac{3}{2 \Omega ^{'2}} \Omega _{\bar{\varphi   }}, \,\,
K_{\varphi   \bar{\lambda   }} \,=\, - \frac{3}{2 \Omega ^{'2}} \Omega _{\varphi   },
\]
and the scalar potential which, for the specific linear dependence of $\Omega '$ on $\lambda $ and $ \bar{\lambda} $, takes on the following form
\[
\begin{array}{ccc}
V \,&=&\, \displaystyle \frac{9}{\Omega ^{'2}} \Big\{ \frac{1}{\Omega _{\varphi  \bar{\varphi }}} \left[W_\varphi  \bar{W}_{\bar{\varphi }} \,+\, 
2\Omega _\varphi W_\lambda   \bar{W}_{\bar{\varphi }} \,+\, 2\Omega _{\bar{\varphi}} W_\varphi    \bar{W}_{\bar{\lambda  }}
\,+\, 4 \left(\Omega _\varphi \Omega _{\bar{\varphi}} - \Omega ' \Omega _{\varphi  \bar{\varphi }}  \right) W_\lambda \bar{W}_{\bar{\lambda  }}  \right] \\
&&  \hspace*{-6.5cm} \,\, \displaystyle  - \, 6  \, \left[W_\lambda \bar{W} + \bar{W}_{\bar{\lambda  }} W  \right] \Big\}
\, .
\end{array}
\]
For its derivation we have used the fact that 
$\Omega '_\varphi = \Omega _\varphi , \,\,\Omega '_{\bar{\varphi}} = \Omega _{\bar{\varphi}}$ and $\Omega '_\lambda = 
\Omega '_{\bar{\lambda }} = -1/2$. 
Moreover we shall assume that 
\[
\Omega ' \,=\, -3 + \varphi \bar{\varphi } - \zeta (\varphi \bar{\varphi })^2 - \frac{1}{2}(\lambda + \bar{\lambda }) 
\]
or in other words the function $\Omega^\prime$ includes quadratic in $ \varphi \bar{\varphi} $  terms, specified by the constant $\zeta$.
Omitting the quadratic in $\,  \varphi \bar{\varphi }$ term this is of the no-scale type \cite{lahanas}. 
Such terms are  introduced for the stabilization of the potential in the $\varphi $ direction and have been also considered in \cite{su3}. 
For the superpotential we take $ P(\phi, \lambda) = \alpha $, that is a constant,  and also assume a nonlinear function of $g(\Lambda)$ which includes quadratic in $\lambda$ terms. We shall see that these terms may play an important role and their presence upsets the cosmological inflation  scenario. With these in mind the superpotential has the form 
\[
W(\Phi , \Lambda ) \,=\, g(\Lambda ) \, \Phi + \alpha , \quad g(\Lambda ) \,=\,d + d_1 \Lambda + d_2 \Lambda ^2,
\]
and thus  it is described by four arbitrary constants namely $ d, d_1, d_2$ and $\alpha$ .  Note that throughout this paper  the  reduced Planck mass, $\, M_P \equiv {M_{Planck}}/{\sqrt{8 \pi}} $, is set to unity, $ M_P =1 $. 

\begin{figure}[h]
    \centering
    \includegraphics[width=0.5 \textwidth]{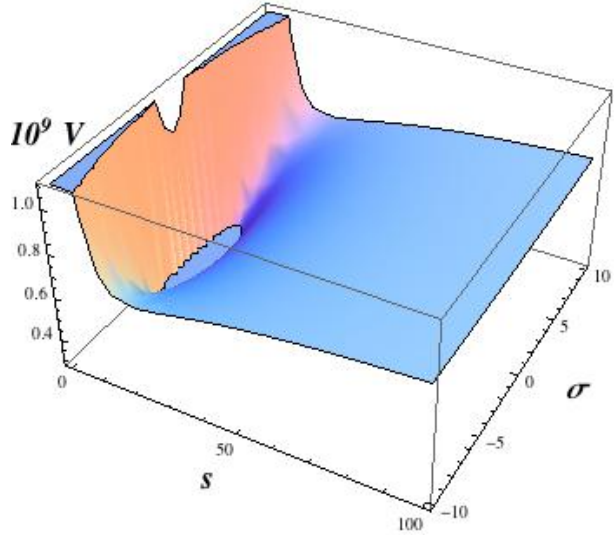} 
    \caption{
    3D plot of the scalar potential of the Lagrangian (\ref{spot})  for values of parameters $ d = -3 \times 10^{-5} , \, d_1 = 10^{-5} ,$ and $ d_2 = 5 \times 10^{-10} $.  The axes are along the real  and the imaginary direction of the field $\lambda \equiv s + i \,  \sigma$ .
     }
    \label{potential}
\end{figure}

At this point it would be useful to make contact with the findings of other authors who use the following notation for the K\"{a}hler function and the super potential \cite{su3},
\bea
 K = - 3 \, ln \, ( T + \bar{T} - C \, \bar{C} \, ) \,  , \, W = 3 M \, C \, ( T - 1 )
 \, .
\eea
Our model corresponds precisely to this model with the identifications
\[
\Lambda = 6 \, T - 3 \,  , \, \Phi = \sqrt{3} \, C 
\]
if  the constants are taken as  
\[
\alpha = d_2 = 0 \, ,  \,d = - 3 d_1 \equiv  
- \frac{\sqrt{3}}{2} \, M 
\]
and the superpotential part $P(\Phi, \Lambda)$ is taken zero. 
In the model under consideration  for simplicity we take $ \alpha = 0 $, in which case there is a minimum at $ \varphi=0$ and the potential is concave in the $ \varphi $ direction. 
Denoting the real and imaginary parts of $ \lambda $ by  $ s $ and  $ \sigma $ respectively the relevant Lagrangian part takes on the form 
\begin{eqnarray}
e^{-1} \mathcal{L}&=& -\frac{1}{2} R -\frac{3}{4}\frac{ \partial_{\mu}{s} \partial^{\mu}{s}}{{ (s+3)^2}} - \frac{3}{4} \frac{ \partial_{\mu}{\sigma} \partial^{\mu}{\sigma}}{{ (s+3)^2}} - \frac{9 \, {| g(s + i \sigma)|}^2}{(s+3)^2}  
\label{spot}
\end{eqnarray} 
where the last part of it is the scalar  potential.  Its form, as function of $s$ and $\sigma$ is displayed in 
Figure \ref{potential}. 
We can see,  by taking  $ d, d_1 $ and $ d_2 $ real, that the minimum of the potential is at $ \sigma=0$.  Ignoring then the fluctuations  of the field $ \sigma $ around its minimum \cite{su13} we have a single-field theory specified by the Lagrangian 
\begin{equation} e^{-1}\mathcal{L} =  -\frac{1}{2} R-\frac{3}{4}\frac{ \partial_{\mu}{s} \partial^{\mu}{s}}{{ (s+3)^2}} -  
\frac{9 \, {| g(s)|}^2}{(s+3)^2}
 \end{equation} 
 whose kinetic terms can appear canonical if we properly redefine the field $s$. 
In particular we   define $\phi$, not to be confused with the scalar component of the superfield $\Phi$ used so far, so that 
\bea
s \, = \, -3 + ( 3+\ell) \, e^{\sqrt{\frac{2}{3}} \, \phi}
\eea 
with the constant $\ell$ defined by
\bea
\ell \, = \, \frac{- d_1 + \sqrt{d_1^2 - 4 \, d \, d_2  }}{2 \, d_2}
\, .
\eea
With these definitions the field $\phi$ is canonically normalized and the potential receives a form reminiscent of the Starobinsky potential. In fact the potential as function of $\phi$ is given by 
\bea
V(\phi) \, = \, 
\frac{3 M^2}{4}    \, { ( \,  1 - e^{- \sqrt{\frac{2}{3}} \, \phi} \, )   }^2  \, 
{|  1 + A \, ( e^{\sqrt{\frac{2}{3}} \phi}  - 1 )   |}^2
\, .
\label{poten}
\eea 
In the expression above the constants  $M, A$ are related to those appearing in the superpotential by 
\bea
M^2 = 12 \, { | d_1 |  }^2 \,( 1- 4 a b )  \; , \;  A = \frac{ 6 b - 1 + \sqrt{ 1- 4 a b } }{2 \, \sqrt{ 1- 4 a b }  }
\eea
where  $a, b$ are the ratios
\[
a = \frac{d}{d_1 } \; , \;  b =  \frac{d_2}{d_1 }
\, .
\]
The parameter $M$ defined above sets  the scale of inflation
\footnote{
As we shall see for small $A$ and moderate $\phi$ the potential (\ref{poten}) is that of Starobinsky having  a plateau where slow-roll can be realised. The normalization of  CMB anisotropies yields then $\, M \simeq 10^{-5}$. 
}

\begin{figure}
    \centering
    \includegraphics[width=0.45 \textwidth]{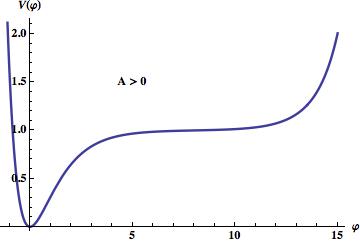} 
    \hspace*{0.1cm}
     \includegraphics[width=0.45 \textwidth]{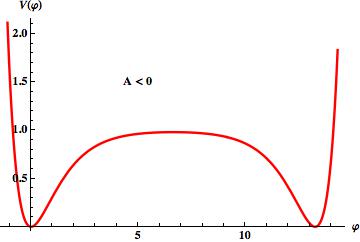} 
         \caption{The general form of the scalar potential (\ref{poten}) as function of $\phi$ for the cases $A > 0 $ ( left panel ) and  
         $A < 0 $ ( right panel )}
    \label{pote}
\end{figure}

Therefore the model is described by the three parameters $d_1, a, b$ but essentially only the combinations  $M, A$ enter into the potential and are relevant for cosmological considerations. 
Note that the case $d_2 = 0$, that is when the quadratic in 
$\Lambda$ terms are absent in $g(\Lambda)$,  corresponds to $A = 0$ and the potential receives the well-known form of the Starobinsky potential.   In that limiting case the value of $\ell$ is $ \ell = - a$, as can be shown by expanding $ \sqrt{ 1- 4 a b } = 1 -2 a b $.  The potential  of Eq. (\ref{poten}) has a single minimum when $A > 0$,  but exhibits two minima when $A < 0$ as shown in figure \ref{pote}.  
In the following we shall consider the $ A > 0 $ case for which the potential exhibits  an almost flat region (plateau)  for values of $\phi$ that are smaller than 
\bea
\phi_i = \sqrt{ \frac{3}{2}  } \, ln \left( 1 + \frac{1}{A} \right)
\, .
\label{phicrit}
\eea
For larger values of $\phi$ the terms proportional to $A$ in the potential dominate and the potential departs from its Starobinsky form exhibiting a rapid exponential behaviour, 
\bea
V(\phi) \, = \, 
\frac{3 M^2}{4}    \, A^2 \,   e^{\, 2 \, \sqrt{\,\frac{2}{3}} \phi} 
\, .
\nonumber
\eea
The coefficient of $\phi$ in the exponent is too large to sustain successful inflation  unless the value of $A$ is small in which case there is a rather  extended  plateau for  values of $\phi $  less than $ \phi_i $.  Then in this region the potential resembles that of Starobinsky's and the inflaton, after a short time,  falls into  the plateau and slow-roll inflation starts. These qualitative features will be quantified numerically in the following sections. In particular we shall see that the features of the Starobinsky inflation are maintained if $A$ is taken at per thousand level so that we have sufficient number of e-foldings, larger than $\sim 60$, for values of $\phi$ belonging to the almost flat regime. This guarantees that in the slow-roll regime a pivot scale $\phi^*$ can be obtained for which the number of e-foldings left, $\, N( \phi^*)  $,  is in the range $50 - 60$ as required.  We have verified, by solving the pertinent equations  numerically, that for typical initial values of the inflaton field,   its value  rapidly drops to $\phi_i$ , following  then a  slow-roll inflation a la  Starobinsky and  the motion of $\phi$ from its initial position to  $\phi_i $ it actually plays little role.  
This will be shown in the following section where a numerical solution is presented.

\section{Numerical treatment of the cosmological equations}

In this section we shall quantify the statements claimed in the previous section by solving the cosmological equation numerically. The system of the pertinent differential equations are well known given by 
\bea
&& \ddot{\phi} + 3 \, H \, \dot{\phi} + V^\prime (\phi) \, = \, 0 \nonumber \\
&& 3 \, H^2 \, = \, \dfrac{{\dot{\phi}}^2}{2} +V(\phi) 
\label{fried}
\eea
where $\, H = \dot{a} / a $ is the Hubble expansion rate. 
We solve (\ref{fried})  for typical initial values of the inflation field at the start of inflation $t=0$.  The cosmic scale factor we take 
$a(t=0) = 1$. Note that since the  potential grows exponentially, with increasing $\phi$, it approaches the Planck energy density  scale for inflaton values  $\phi_P$ dictated by, ( see for instance \cite{Linde:2007fr} ) ,
\bea
V(\phi_P) = 1
\eea
which  entails to 
\bea
\phi_P \, = \, \sqrt{\dfrac{3}{8}} \, ln \left(  \dfrac{4}{3 \, A^2 \, M^ 2} \right) \, \simeq \, 14.28 - 1.22 \, ln \, A
\, .
\eea
The scale of the  inflation $M$ has been set equal to $ M = 10^{-5}$. 
Therefore  typical  initial values at the beginning of inflation  are  given by 
\bea
\phi \sim \phi_P \quad  \,  \quad {\dot{\phi_P}}^2 / 2 \sim V(\phi_P) = 1
\, .
\eea
For $A=1$ , $\phi_P \sim 14 $ while for small $ A = 10^{-4}$ the initial inflation value is  $\phi_P \sim 25 $. 

\begin{figure}[h]
    \centering
    \includegraphics[width=0.45 \textwidth]{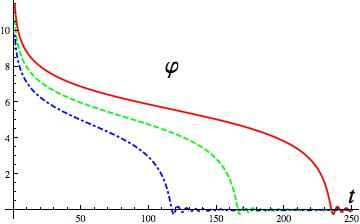} 
    \hspace*{0.1cm}
     \includegraphics[width=0.45 \textwidth]{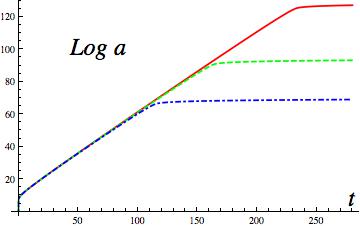} 
\caption{
Evolution of the inflaton $\phi$ (left) and the logarithm of the  cosmic scale factor $log a$ (right)  with time. The time is taken in units of the scale  $M = 10^{-5}$. The solid (Red) , dashed (Green) and dash - dot (Blue) lines correspond to values of $A = \,10^{-4}, 2 \times 10^{-4},   4  \times 10^{-4}  $ respectively.  
}
  \label{evo1}
\end{figure}

\begin{figure}[h]
    \centering
    \includegraphics[width=0.45 \textwidth]{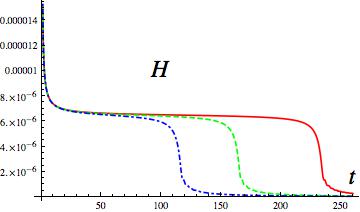} 
    \hspace*{0.1cm}
     \includegraphics[width=0.45 \textwidth]{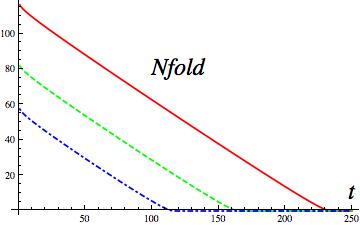} 
 \caption
 {Evolution of the Hubble rate $H$  (left) and the number of e-foldings $\mathrm{\it{Nfold}}$ (right)  with time.  $M, A$ are as in figure \ref{evo1}. 
 }
    \label{evo2}
\end{figure}

The evolution for the inflaton $\phi$ and the cosmic scale factor for various  values of the parameter $A$ are displayed in Figure \ref{evo1} . The solid (Red), dashed (Green) and dash - dot (Blue)  lines correspond to  $\, A = 10^{-4},  \, 2 \times 10^{-4} \,$ and 
  $\, A = 4 \times 10^{-4} \,$ respectively. 
From the left panel of this figure we see that after a sharp drop the inflaton follows a normal slow-roll evolution during which the Universe undergoes a de-Sitter expansion as  shown on the right panel of this figure where the evolution of the cosmic scale factor is shown. The horizontal axis is the time  in units of the inflation scale $M = 10^{-5}$. The exit from inflation occurs when $\, log a$ starts becoming almost constant.  Note the damped oscillatory behaviour of $\, \phi$ as it drops within the minimum of the potential.  
In Figure \ref{evo2}, on the left panel, the evolution of the Hubble rate is shown while on the right we show the number of 
e-foldings as function of time,  $ \, N(t) \equiv N(\phi(t))$.  
The parameters are as in Figure \ref{evo1}. One observes  that 
after a short drop-off $H$ enters into the slow-roll era, during which it  stays almost constant,  exiting from it at 
a time, in each case displayed, that coincides with the time signalling departure of $H$ from its constancy, as expected. 
The number of e-foldings  $\, N(t) $  should be in the range 
is  $\, \sim 50 - 60$ at a time $\, t^*   $ at which the inflation receives the pivot  value $\, \phi^* $. One notices that for the 
largest of the  sample values chosen, $A = 4 \times10^{-4}$,  we can marginally obtain a  pivot value for which the number of the e-foldings left is in the aforementioned range. As a conclusion only small values of the $A$ parameter  are allowed which we have found to be bounded by $\, A \leq 5 \times 10^{-4}$. 
Note that for  moderate to large values of the parameter $A$ the potential is mainly exponential $\, \sim e^{\, \lambda \, \phi}$, with 
$\,  \lambda = 2 \sqrt{2 / 3}$, and such large values of $\, \lambda$  in exponential potentials are known to be  incompatible with slow-roll, unless $\, \lambda^2 < 2$. 
Therefore the exponential part by itself cannot sustain inflation and an extended plateau is required, on which the inflaton rolls  after a rapid fall from the exponential region.  This plateau  must be extended enough to obtain the desired number of e-foldings which is quantified by the above upper bound on $\, A$.  

 Although this inflationary scenario resembles that of Starobinsky's model  nevertheless  both  the slow-roll parameters $\epsilon, \eta$ depend on the additional parameter $A$ which although small may  affect the value of 
$\epsilon$ and hence the tensor to scalar ratio $r$.  Whether large values for $r$ can be obtained and if these can be consistent with  the remaining observables, in particular  the spectral index $n_s$  will be investigated in the following section. 

\section{Slow-roll inflation}

From the discussion of the previous section it becomes evident that only small values of $A$ are allowed and all cosmological parameters depend on it. 
However small this might be it may have, in principle,  a large impact on the slow-roll parameters and the number of 
e-foldings. 

\begin{figure}[h]
    \centering
    \includegraphics[width=0.5 \textwidth]{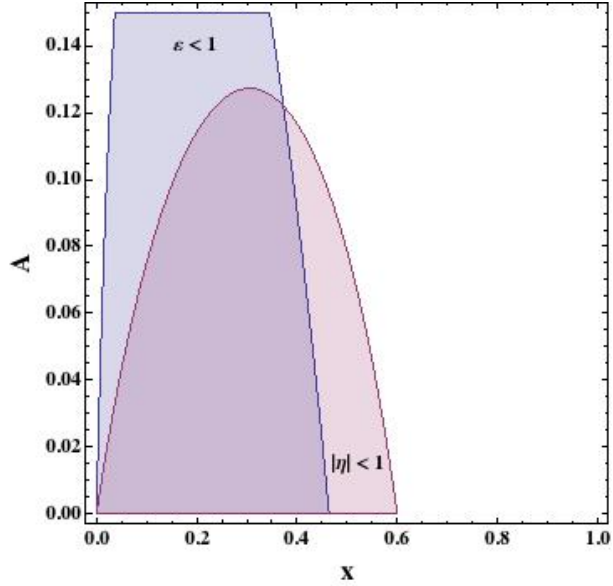} 
    \caption{The shadowed areas designates the region where slow-roll approximation holds.  $ \epsilon < 1 $ ( in Blue ) allows for large $A$ values while  $\eta < 1 $ ( in Magenta ) requires $ A < .13 $.  }
    \label{ereg}
\end{figure}

Starting from the latter,  we have already seen that  it puts an upper bound on the allowed value of $A < 5 \times 10^{-4}$ by the requirement that the number of $N$ left until the exit from inflation is in the range $\, 50 - 60$.  Its analytic form, for any value of $\phi$ in the range 
$\phi < \phi_i$,  can be calculated analytically given by
\bea
N(x) \, = \, - \frac{3}{4} \, ln \left[  \frac{\; \;  x \; \;( k + x_{end}^2 )}{x_{end} \; (k + x^2 )} \right]  + \frac{3 ( 1 - k ) }{4 \, \sqrt{k}} \, 
\left(  arctan\frac{x_{end}}{\sqrt{k}}  -  arctan\frac{x}{\sqrt{k}}  \right)
\, .
\label{NNN}
\eea
In this $x$ stands for the more convenient variable $\, x = \exp \left( -\sqrt{\frac{3}{2}} \, \phi \right)$ through which the region 
$\, \phi= [ 0, +\infty ]$ is mapped  to $\, x = [1, 0]$. The constant $k$ is $ k \equiv \frac{A}{1-A}$. The equation (\ref{NNN})  holds for any  $A < 1$ which is always the case when $A$ is small.  The value of $x_{end}$ signalling the end of the inflation period  is determined when at least one or both of $\epsilon, \eta $ become of order unity. In Figure \ref{ereg} we display the regions where slow-roll conditions hold  in the $\, x, A$ plane.  The slow-roll region, where both $\epsilon, \eta $  are less than unity, allows for values 
$\, A < 0.13$.  For each $A$ the furthest point on the right of the allowed overlapping region marks the end of the inflation point  
$\, x_{end}$. For small $A$ this is given analytically, to a good approximation, by
\bea
x_{end} \equiv e^{ - \sqrt{\frac{2}{3}} \, \phi_{end}  }  \simeq \ {\sqrt{3}}/ ({2+ \sqrt{3}}  ) \, ( 1 - A ) \simeq 0.5 \, ( 1 - A )
\, .
\eea 
Note that the above upper bound, $A < 0.13$,  is much larger than the bound set by the requirement to have sufficient number of  
e-foldings   and therefore for values  $ A < 5 \times 10^{-4}$  we are well within the slow-roll regime. 

Concerning the slow roll parameter $\, \epsilon$ and in order to study its sensitivity on the parameter $A$ it facilitates if we write potential of Eq. ( \ref{poten})  as
\bea
V(\phi) = V_S (\phi) \, f(\phi)
\label{potfa}
\eea
where $\, V_S$ is the Starobinsky potential and $f(\phi)$ the additional factor that dominates for large values of $\phi$ and depends on the parameter $A$. Then in a straightforward manner one can show that the slow-roll parameter $\epsilon$, as function of the inflaton field, or the field $\, x$ defined before, is related to the corresponding value found for the Starobinsky potential, denoted by 
$\epsilon_s$, by 
\bea
\sqrt{\epsilon} \, = \, \sqrt{\epsilon_s} \, + \, \frac{2}{\sqrt{3}} \, \frac{  1  }{  1 + ( A^{-1} - 1 ) \,x \,   }   
\, .
\label{epsit}
\eea
One observes that 
for the same value of the inflation field  the $\, \epsilon$ parameter in the model studied here is larger than 
$\,  \epsilon_s $ by amounts that are controlled by the parameter $A$ as is evident from Eq. (\ref{epsit}) .  This equation is valid provided we  are in the slow-roll regime which entails to values of $\, \phi$  for which the additional term, appearing on the right hand side of Eq (\ref{epsit}), is much less than unity, 
\bea
\phi << \sqrt{  \frac{3}{2}  } \, ln ( A^{-1} -1 ) 
\, . 
\label{eneq}
\eea
Therefore slow roll in the Starobinsky model yields slow-roll motion in the model under consideration in the regime specified by the inequality above. Note that since $A$ has to be small the rhs of the equation above is very close to $\, \phi_i$ which signals departure from the simple Starobinsky potential. Although small in the regime of interest the additional term in (\ref{epsit}) may not be small as compared to $\, \sqrt{\epsilon_s} $ and may significantly augment the value of the epsilon parameter tending to increase the ratio $\, r$ of the tensor to scalar perturbations. The precise amount depends on the parameter $\, A$ but the analysis has to take into account the remaining cosmological data, in particular the number of e-foldings, which specifies the pivot scale, and the value of the spectral index $\,  \eta_s$ as well. 

\begin{figure}[h]
    \centering
    \includegraphics[width=0.6 \textwidth]{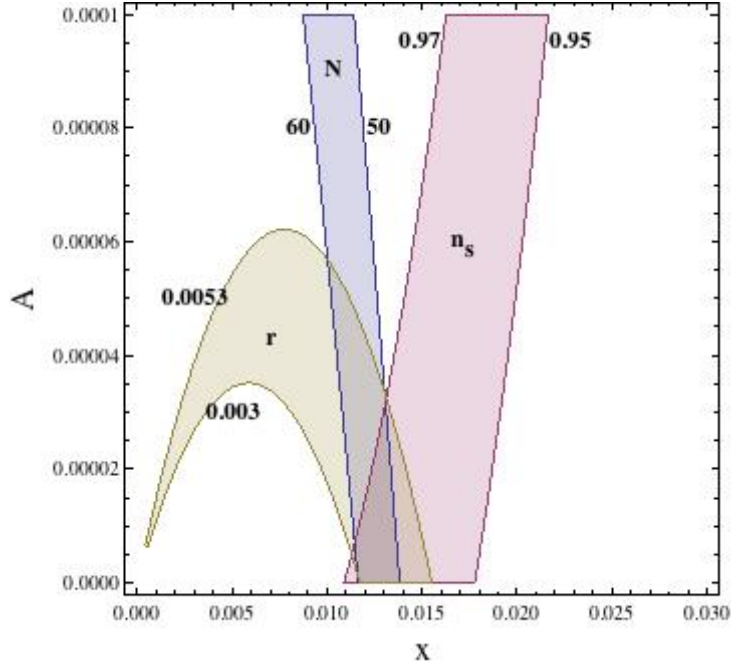} 
    \caption{Regions for $N, r, n_s$ within the ranges  designated at the borders of each region.  The allowed   $r$  region by $N$ and $n_s$  predicts $r$ in the range $0.003 < r < 0.0053 $ although the $N$ and $n_s$ separately each  allows for much higher values.   }
    \label{rnn}
\end{figure}

In Figure \ref{rnn} we plot contour values for the ratio  $\, r$ the index $\,  \eta_s$ and the number of e-foldings, as functions of the field variable $\, x$. On the vertical axis are the values of the parameter $A$. 
Large values of $\, r$ are allowed by $N$ and $n_s$ observational constraints separately ( not shown in Figure \ref{rnn}).  However the intersection of $N$ and $n_s$ restricts the allowed region to values of $A$ that are smaller than $ 4.0 \, \times \,10^{-5}$ and values of $x$ that are around $0.013$ as shown in the figure. Within this region the allowed values of $r$ are $0.003< r < 0.0053$. 
Note that the displayed region which is allowed  by all data is well within the slow-roll regime. 
Therefore,  although the model  can yield larger values of $r$,  as compared to the simple Starobinsky model, which are  
in agreement with the new data released by Planck and BICEP2,  nevertheless  it rather favours low values that cannot saturate the upper bound imposed on $r$ by these experiments. 

Concluding this section,  the inclusion of quadratic in $\Lambda$ terms in the superpotential  introduces a new parameter in the potential that can result, in principle, to larger $r$ values, in comparison with the linear ( in $\Lambda$ ) supersymmetric Starobinsky model, but not sufficiently large to  approach values of $r$ close to  the upper  bound set by recent data of  Planck and BICEP2
if all observational constraints are taken into account. 
The fact that one needs very small $A$ values to maintain the good features of the Starobinsky's model, at least as far as the Planck satellite data are concerned, indicates that  non-linear superpotential $\, \Lambda$ terms are only allowed provided that they have small couplings.

\section{Discussion - Conclusions}
In this note we generalize the supergravity Starobinsky models allowing for superpotential terms that are not linear in the superfield 
 $\, \Lambda$ that couples to the chiral Ricci multiplet.  We exemplify the departure from the linearity by the 
 the appearance of quadratic in this field terms. The K\"ahler function is assumed to be of the no-scale type.  
 The inclusion of such terms results to introducing additional parameters  in the theory which however enter into the scalar potential in two combinations. One, $M$,  sets the scale of the inflationary potential and the other, $A$, deforms the Starobinsky potential in a multiplicative way. In the limit $A=0$, in which case the quadratic terms are absent,  one recovers the simple Starobinsky model but for $A \neq 0$ the potential deviates from it increasing exponentially for large inflaton values. Successful inflation is achieved only if the parameter $A$ is smaller than $\sim 5 \times 10^{-4}$, for natural initial values of the inflaton field. This constraint stems mainly from the requirement to have sufficient  number of e-foldings  left to the end of inflation. For such  small values of $A$ the inflaton starts its motion dropping  rapidly to the Starobinsky plateau which is rather extended due to the smallness of the parameter $A$. 
The dependence of the slow-roll parameters $\, \epsilon$, and hence of $r$,  on the additional parameter $A$  is investigated. Although in principle the tensor to scalar ratio can be theoretically much larger in comparison with the  predictions of the Starobinsky model, nevertheless the imposition of the data concerning the spectral index $n_s$ in combination with the required   number of e-foldings suppresses the allowed  values of $A$ even more, by almost an order of magnitude, in regions where $r$ cannot exceed  
$ r \simeq 0.005$. 
This value is slightly larger than the one predicted in the  Starobinsky model but in no case can saturate the new combined bound released  by Planck and BICEP2 experiments if all data are observed. 
On the theoretical side, the smallness of the parameter $A$, as a result of the comparison with the observational data,  indicates that the presence of non-linear in $\, \Lambda$  terms within the superpotential is only possible  provided  that these are  very small. This fine tuning  suggests  that the departure from the non-linearity is presumably due to quantum effects.

\vspace*{6mm}
\noindent
{\textbf{Acknowledgements}}

This research has been co-financed by the European Union (European Social Fund - ESF)
and Greek national funds through the Operational Program "Education and Lifelong
Learning" of the National Strategic Reference Framework (NSRF) - Research Funding
Program: {\textit{"THALES"-Investing in the society of knowledge through the European Social Fund.}}  
A.B. L would also like to thank CERN Theory Division for the hospitality where this work was completed 
and K. Papadodimas for discussions. A.B.L. would also like to thank K. Tamvakis for discussions and his critical remarks 
concerning the content of this work.


\end{document}